# Mutual derivation between arbitrary distribution forms of momenta and momentum components

Pei-Pin Yang, Qi Wang, Fu-Hu Liu*

*Institute of Theoretical Physics & State Key Laboratory of Quantum Optics and Quantum Optics Devices,*

*Shanxi University, Taiyuan, Shanxi 030006, China*

* E-mail: fuhuliu@163.com; fuhuliu@sxu.edu.cn

**Abstract:** The mutual derivation between arbitrary distribution forms of momenta and momentum components of particles produced in an isotropic emission source are systematically studied in terms of probability theory and mathematical statistics. The distributions of rapidities and pseudorapidities are expediently studied. As an example, the classical and relativistic ideal gas models are used to show these distributions by the analytic and Monte Carlo methods. As an application, the experimental rapidity and transverse momentum spectra of light flavor particles produced in high energy collisions are analyzed by a multi-component relativistic ideal gas model in which the single model can be replaced by other models and distributions.

## 1. Introduction

In high energy hadron-hadron [1-3], hadron-nucleus [4, 5], and nucleus-nucleus collisions [6-10], some distributions (probability density functions) such as the distributions of transverse momenta and (pseudo)rapidities can be measured in experiments [11, 12]. This is convenient for researchers to study these distributions in theory. However, because of the limitation of experimental conditions, other distributions cannot be measured directly or indirectly in experiments. Instead, one can derive other distributions from the measured distributions based on the probability theory and mathematical statistics [13]. In particular, although some distributions look outwardly to be simple and easy, there are few systematical studies on the relationships among them.

Not only in the field of high energy collisions, but also in other fields related to probability, statistics, and data science, the mutual derivation between different distributions are needed in some cases. In addition, the distributions of rapidities and pseudorapidities have been treated as approximate equivalent in many cases. We are interested in the degree of their coincidences, though the two distributions are known to be approximately the same at high energy or high transverse momentum. Based on the invariant momentum distribution, the distributions of transverse momenta and rapidities are related to be each other. Their joint distribution and respective distributions can be obtained from the invariant momentum distribution.

To understand the relations among different distributions, the mutual derivation between arbitrary distribution forms of momenta and momentum components of particles produced in an isotropic emission source are systematically studied in this paper. Meanwhile, the distributions of rapidities and pseudorapidities are expediently studied. The classical [14] and relativistic ideal gas models [15, 16] are used as examples to show these distributions by the analytic and Monte Carlo methods so that we can test and verify their results each other. As an application, the experimental rapidity and transverse momentum spectra of negative pions produced in high energy collisions are analyzed by a multi-component relativistic ideal gas model. In this study, the most basic consideration is the ideal gas model which was considered the first time by L. D. Landau in his outstanding paper in 1953 [17].

The structure of this paper is in the following. Sections 2 and 3 describe the method of mutual derivation among different distributions. Section 4 shows some distributions as examples by the analytic and Monte Carlo methods based on the ideal gas model, or rapidity and transverse momentum spectra of light flavor particles produced in high energy collisions analyzed by a multi-component relativistic ideal gas model. Finally, the conclusion is summarized in section 5.

**2. Distributions derived from distributions of momentum components**

In the case of the distributions of momentum components $p_x$, $p_y$, and $p_z$ being known, other distributions such as the distributions of transverse momenta $p_T$, momenta $p$, azimuth angles $\phi$, and emission angles $\theta$ can be obtained. Let $f_x(x) = (1/N) dN/dx$ denote the distribution of variable $x$, where $x$ denotes one of the above mentioned quantities and $N$ denotes the number of particles. Generally, in the rest frame of an emission source, we can assume the emission of particles to be isotropic in the three-dimensional momentum space. At the same time, we assume that $f_{p_x}(p_x)$, $f_{p_y}(p_y)$, and $f_{p_z}(p_z)$ are independent, $f_{p_T}(p_T)$ and $f_\phi(\phi)$ are independent, and $f_p(p)$ and $f_\theta(\theta)$ are independent.

According to the relations $p_T = \sqrt{p_x^2 + p_y^2}$, $p_x = p_T \cos\phi$, $p_y = p_T \sin\phi$, and $dp_x dp_y = p_T dp_T d\phi$, we have the relation between the joint distribution $f_{p_x, p_y}(p_x, p_y)$ of $p_x$ and $p_y$ and the joint distribution $f_{p_T, \phi}(p_T, \phi)$ of $p_T$ and $\phi$ to be [13]

$$f_{p_x, p_y}(p_x, p_y) |dp_x dp_y| = f_{p_T, \phi}(p_T, \phi) |dp_T d\phi|. \tag{1}$$

Then,

$$\begin{aligned} f_{p_T, \phi}(p_T, \phi) &= p_T f_{p_x, p_y}(p_x, p_y) = p_T f_{p_x}(p_x) f_{p_y}(p_y) \\ &= p_T f_{p_x}(p_T \cos\phi) f_{p_y}(p_T \sin\phi) \end{aligned}, \tag{2}$$

$$f_{p_T}(p_T) = \int_0^{2\pi} f_{p_T, \phi}(p_T, \phi) d\phi = p_T \int_0^{2\pi} f_{p_x}(p_T \cos\phi) f_{p_y}(p_T \sin\phi) d\phi, \tag{3}$$

$$f_\phi(\phi) = \int_0^\infty f_{p_T, \phi}(p_T, \phi) dp_T = \int_0^\infty p_T f_{p_x}(p_T \cos\phi) f_{p_y}(p_T \sin\phi) dp_T. \tag{4}$$

In addition, for the isotropic emission process and based on the consideration of normalization, we have

$$f_\phi(\phi) = \frac{1}{2\pi}, \tag{5}$$

which is an even distribution in $[0, 2\pi]$. In particular, in the classical ideal gas model [14],

$$f_{p_x}(p_x) = \frac{1}{\sqrt{2\pi}\sigma} \exp\left(-\frac{p_x^2}{2\sigma^2}\right), \tag{6}$$

$$f_{p_y}(p_y) = \frac{1}{\sqrt{2\pi}\sigma} \exp\left(-\frac{p_y^2}{2\sigma^2}\right), \tag{7}$$

$$f_{p_z}(p_z) = \frac{1}{\sqrt{2\pi}\sigma} \exp\left(-\frac{p_z^2}{2\sigma^2}\right), \tag{8}$$

where $\sigma = \sqrt{m_0 T}$ denotes the distribution width, $m_0$ denotes the rest mass of considered particle, and $T$ denotes the temperature of emission source. We have

$$f_{p_T}(p_T) = \frac{p_T}{2\pi\sigma^2} \exp\left(-\frac{p_T^2}{2\sigma^2}\right) \int_0^{2\pi} d\phi = \frac{p_T}{\sigma^2} \exp\left(-\frac{p_T^2}{2\sigma^2}\right), \tag{9}$$

which is the Rayleigh distribution in $[0, \infty)$. $f_\phi(\phi)$ is shown in Eq. (5).

According to the relations $p = \sqrt{p_x^2 + p_y^2 + p_z^2}$, $p_x = p\sin\theta\cos\phi$, $p_y = p\sin\theta\sin\phi$, $p_z = p\cos\theta$, and $dp_x dp_y dp_z = p^2 \sin\theta dp d\theta d\phi$, we have the relation between the joint distribution $f_{p_x,p_y,p_z}(p_x,p_y,p_z)$ of $p_x$, $p_y$, and $p_z$ as well as the joint distribution $f_{p,\theta,\phi}(p,\theta,\phi)$ of $p$, $\theta$, and $\phi$ to be [13]

$$f_{p_x,p_y,p_z}(p_x,p_y,p_z)|dp_x dp_y dp_z| = f_{p,\theta,\phi}(p,\theta,\phi)|dp d\theta d\phi|. \tag{10}$$

Then,

$$\begin{aligned} f_{p,\theta,\phi}(p,\theta,\phi) &= p^2 \sin\theta f_{p_x,p_y,p_z}(p_x,p_y,p_z) \\ &= p^2 \sin\theta f_{p_x}(p_x) f_{p_y}(p_y) f_{p_z}(p_z) \\ &= p^2 \sin\theta f_{p_x}(p\sin\theta\cos\phi) f_{p_y}(p\sin\theta\sin\phi) f_{p_z}(p\cos\theta) \end{aligned} \tag{11}$$

$$\begin{aligned} f_p(p) &= \int_0^\pi \int_0^{2\pi} f_{p,\theta,\phi}(p,\theta,\phi) d\phi d\theta \\ &= p^2 \int_0^\pi \int_0^{2\pi} \sin\theta f_{p_x}(p\sin\theta\cos\phi) f_{p_y}(p\sin\theta\sin\phi) f_{p_z}(p\cos\theta) d\phi d\theta \end{aligned} \tag{12}$$

$$\begin{aligned} f_\theta(\theta) &= \int_0^\infty \int_0^{2\pi} f_{p,\theta,\phi}(p,\theta,\phi) d\phi dp \\ &= \sin\theta \int_0^\infty \int_0^{2\pi} p^2 f_{p_x}(p\sin\theta\cos\phi) f_{p_y}(p\sin\theta\sin\phi) f_{p_z}(p\cos\theta) d\phi dp \end{aligned} \tag{13}$$

In addition, for the isotropic emission process and based on the consideration of normalization, we have

$$f_\theta(\theta) = \frac{1}{2}\sin\theta, \tag{14}$$

which is a half sine distribution in $[0,\pi]$. In particular, in the classical ideal gas model [14],

$$f_p(p) = \frac{p^2}{(2\pi)^{3/2}\sigma^3} \exp\left(-\frac{p^2}{2\sigma^2}\right) \int_0^\pi \int_0^{2\pi} \sin\theta d\phi d\theta = \sqrt{\frac{2}{\pi}} \frac{p^2}{\sigma^3} \exp\left(-\frac{p^2}{2\sigma^2}\right), \tag{15}$$

which is the Maxwell distribution in $[0,\infty)$. $f_\theta(\theta)$ is shown in Eq. (14).

According to the relations $p = \sqrt{p_T^2 + p_z^2}$, $p_T = p\sin\theta$, $p_z = p\cos\theta$, and $dp_T dp_z = pdp d\theta$, we have the relation between the joint distribution $f_{p_T,p_z}(p_T,p_z)$ of $p_T$ and $p_z$ as well as the joint distribution $f_{p,\theta}(p,\theta)$ of $p$ and $\theta$ to be [13]

$$f_{p_T,p_z}(p_T,p_z)|dp_T dp_z| = f_{p,\theta}(p,\theta)|dp d\theta|. \tag{16}$$

Then,

$$f_{p,\theta}(p,\theta) = p f_{p_T,p_z}(p_T, p_z) = p f_{p_T}(p_T) f_{p_z}(p_z) = p f_{p_T}(p\sin\theta) f_{p_z}(p\cos\theta), \quad (17)$$

$$f_p(p) = \int_0^\pi f_{p,\theta}(p,\theta) d\theta = p\int_0^\pi f_{p_T}(p\sin\theta) f_{p_z}(p\cos\theta) d\theta, \quad (18)$$

$$f_\theta(\theta) = \int_0^\infty f_{p,\theta}(p,\theta) dp = \int_0^\infty p f_{p_T}(p\sin\theta) f_{p_z}(p\cos\theta) dp. \quad (19)$$

In particular, for the isotropic emission process and in the classical ideal gas model, $f_p(p)$ and $f_\theta(\theta)$ are shown in Eqs. (15) and (14) respectively.

It is difficult to obtain the rapidity ($y \equiv (1/2)\ln[(E+p_z)/(E-p_z)]$) distribution from the distributions of momentum components, where $E = \sqrt{p^2 + m_0^2} = m_T \cosh y$ is the energy and $m_T = \sqrt{p_T^2 + m_0^2}$ is the transverse mass. Instead, the pseudorapidity ($\eta \equiv -\ln\tan(\theta/2)$) distribution can be obtained due to Eq. (14). We have

$$f_\eta(\eta) = f_\theta(\theta)\left|\frac{d\theta}{d\eta}\right| = \frac{1}{2\cosh^2 \eta}. \quad (19)$$

In some cases, we need the Monte Carlo method to obtain some quantities [18]. The distributions of these quantities are then obtained by statistics. Let $R$, $R_{1,2,3,4,5,6}$, and $r_{1,2,3}$ denote random numbers distributed evenly in $[0,1]$. Generally, $x$ satisfies

$$\int_{x_{\min}}^{x} f_x(x') dx' < R < \int_{x_{\min}}^{x+\delta x} f_x(x') dx', \quad (20)$$

where $x_{\min}$ denotes the minimum $x$ and $\delta x$ denotes a small amount. If the integral can result in a special expression or one has other special expressions, the treatment will be easier.

Concretely, in the classical ideal gas model, we have

$$p_x = \sigma\sqrt{-2\ln R_1}\cos(2\pi R_2), \text{ or } p_x = \sigma\sqrt{-2\ln R_1}\sin(2\pi R_2), \quad (21)$$

$$p_y = \sigma\sqrt{-2\ln R_3}\cos(2\pi R_4), \text{ or } p_y = \sigma\sqrt{-2\ln R_3}\sin(2\pi R_4), \quad (22)$$

$$p_z = \sigma\sqrt{-2\ln R_5}\cos(2\pi R_6), \text{ or } p_z = \sigma\sqrt{-2\ln R_5}\sin(2\pi R_6), \quad (23)$$

$$\phi = 2\pi r_1, \quad (24)$$

$$\theta = 2\arcsin\sqrt{r_2}. \quad (25)$$

Then, other quantities can be obtained by their definitions or relations with the above quantities. In particular, $p_T$ is also given by

$$p_T = \sigma\sqrt{-2\ln r_3}. \quad (26)$$

## 3. Distributions derived from distributions of transverse momenta and momenta

From any $f_{p_T}(p_T)$, we can obtain $f_{p_x}(p_x)$ and $f_{p_y}(p_y)$ due to their independent character and the isotropic assumption. According to Eqs. (1) and (5), we have [13]

$$f_{p_x}(p_x)f_{p_y}(p_y)|dp_xdp_y| = \frac{1}{2\pi p_T}f_{p_T}(p_T)|dp_xdp_y|$$
$$= \frac{1}{2\pi\sqrt{p_x^2+p_y^2}}f_{p_T}\left(\sqrt{p_x^2+p_y^2}\right)|dp_xdp_y| \quad . \tag{27}$$

Then,

$$f_{p_x}(p_x) = \frac{1}{2\pi}\int_{-\infty}^{\infty}\frac{1}{\sqrt{p_x^2+p_y^2}}f_{p_T}\left(\sqrt{p_x^2+p_y^2}\right)dp_y, \tag{28}$$

$$f_{p_y}(p_y) = \frac{1}{2\pi}\int_{-\infty}^{\infty}\frac{1}{\sqrt{p_x^2+p_y^2}}f_{p_T}\left(\sqrt{p_x^2+p_y^2}\right)dp_x. \tag{29}$$

From any $f_p(p)$, we can obtain $f_{p_T}(p_T)$ and $f_{p_z}(p_z)$ due to their independent character and the isotropic assumption in the three-dimensional momentum space. According to Eqs. (16) and (14), we have [13]

$$f_{p_T}(p_T)f_{p_z}(p_z)|dp_Tdp_z| = \frac{\sin\theta}{2p}f_p(p)|dp_Tdp_z|$$
$$= \frac{p_T}{2(p_T^2+p_z^2)}f_p\left(\sqrt{p_T^2+p_z^2}\right)|dp_Tdp_z| \quad . \tag{30}$$

Then,

$$f_{p_T}(p_T) = \frac{p_T}{2}\int_{-\infty}^{\infty}\frac{1}{(p_T^2+p_z^2)}f_p\left(\sqrt{p_T^2+p_z^2}\right)dp_z, \tag{31}$$

$$f_{p_z}(p_z) = \frac{1}{2}\int_{0}^{\infty}\frac{p_T}{(p_T^2+p_z^2)}f_p\left(\sqrt{p_T^2+p_z^2}\right)dp_T. \tag{32}$$

Based on any $f_p(p)$ and the isotropic assumption in the three-dimensional momentum space, we have the invariant momentum distribution

$$E\frac{d^3N}{dp^3} \propto \frac{E}{p^2}f_p(p) = \frac{E}{E^2-m_0^2}f_p\left(\sqrt{E^2-m_0^2}\right). \tag{33}$$

Then,

$$\frac{d^2N}{dydp_T} \propto \frac{p_Tm_T\cosh y}{m_T^2\cosh^2 y - m_0^2}f_p\left(\sqrt{m_T^2\cosh^2 y - m_0^2}\right), \tag{34}$$

$$f_{p_T}(p_T) \propto p_Tm_T\int_{y_{\min}}^{y_{\max}}\frac{\cosh y}{m_T^2\cosh^2 y - m_0^2}f_p\left(\sqrt{m_T^2\cosh^2 y - m_0^2}\right)dy, \tag{35}$$

$$f_y(y) \propto \cosh y \times \int_0^{p_{T\max}}\frac{p_Tm_T}{m_T^2\cosh^2 y - m_0^2}f_p\left(\sqrt{m_T^2\cosh^2 y - m_0^2}\right)dp_T. \tag{36}$$

In particular, in the classical ideal gas model [14],

$$f_p(p) = \sqrt{\frac{2}{\pi}} \frac{p^2}{(m_0 T)^{3/2}} \exp\left(-\frac{p^2}{2m_0 T}\right), \tag{37}$$

$$E\frac{d^3N}{dp^3} \propto E\exp\left(-\frac{p^2}{2m_0 T}\right) = E\exp\left(-\frac{E^2 - m_0^2}{2m_0 T}\right), \tag{38}$$

$$\begin{aligned}\frac{d^2N}{dy\,dp_T} &\propto p_T m_T \cosh y \times \exp\left[-\frac{m_T^2 \cosh^2 y - m_0^2}{2m_0 T}\right] \\ &\propto p_T m_T \cosh y \times \exp\left[-\frac{m_T^2 \cosh^2 y}{2m_0 T}\right]\end{aligned} \tag{39}$$

where the item $\exp[m_0^2/2m_0 T]$ is included in the normalization constant,

$$f_{p_T}(p_T) \propto p_T m_T \int_{y_{\min}}^{y_{\max}} \cosh y \times \exp\left[-\frac{m_T^2 \cosh^2 y}{2m_0 T}\right] dy$$

$$\text{or} \quad f_{p_T}(p_T) = \frac{p_T}{m_0 T} \exp\left(-\frac{p_T^2}{2m_0 T}\right), \tag{40}$$

$$f_y(y) \propto \cosh y \times \int_0^{p_{T\max}} p_T m_T \exp\left[-\frac{m_T^2 \cosh^2 y}{2m_0 T}\right] dp_T, \tag{41}$$

$$f_{p_x}(p_x) \propto \int_{-\infty}^{\infty} \sqrt{p_x^2 + p_y^2 + m_0^2} \int_{y_{\min}}^{y_{\max}} \cosh y \times \exp\left[-\frac{(p_x^2 + p_y^2 + m_0^2)\cosh^2 y}{2m_0 T}\right] dy\,dp_y$$

$$\text{or} \quad f_{p_x}(p_x) = \frac{1}{2\pi m_0 T} \int_{-\infty}^{\infty} \exp\left(-\frac{p_x^2 + p_y^2}{2m_0 T}\right) dp_y = \frac{1}{\sqrt{2\pi m_0 T}} \exp\left(-\frac{p_x^2}{2m_0 T}\right), \tag{42}$$

$$f_{p_y}(p_y) \propto \int_{-\infty}^{\infty} \sqrt{p_x^2 + p_y^2 + m_0^2} \int_{y_{\min}}^{y_{\max}} \cosh y \times \exp\left[-\frac{(p_x^2 + p_y^2 + m_0^2)\cosh^2 y}{2m_0 T}\right] dy\,dp_x$$

$$\text{or} \quad f_{p_y}(p_y) = \frac{1}{2\pi m_0 T} \int_{-\infty}^{\infty} \exp\left(-\frac{p_x^2 + p_y^2}{2m_0 T}\right) dp_x = \frac{1}{\sqrt{2\pi m_0 T}} \exp\left(-\frac{p_y^2}{2m_0 T}\right), \tag{43}$$

$$f_{p_z}(p_z) = \frac{1}{\sqrt{2\pi}(m_0 T)^{3/2}} \int_0^{\infty} p_T \exp\left(-\frac{p_T^2 + p_z^2}{2m_0 T}\right) dp_T = \frac{1}{\sqrt{2\pi m_0 T}} \exp\left(-\frac{p_z^2}{2m_0 T}\right). \tag{44}$$

Meanwhile, in the relativistic ideal gas model [15, 16],

$$f_p(p) \propto p^2 \exp\left(-\frac{\sqrt{p^2 + m_0^2}}{T}\right), \tag{45}$$

$$E\frac{d^3N}{dp^3} \propto E\exp\left(-\frac{\sqrt{p^2 + m_0^2}}{T}\right) = E\exp\left(-\frac{E}{T}\right), \tag{46}$$

$$\frac{d^2N}{dydp_T} \propto p_T m_T \cosh y \times \exp\left[-\frac{m_T \cosh y}{T}\right], \tag{47}$$

$$f_{p_T}(p_T) \propto p_T m_T \int_{y_{\min}}^{y_{\max}} \cosh y \times \exp\left[-\frac{m_T \cosh y}{T}\right] dy,$$

$$\text{or} \quad f_{p_T}(p_T) \propto p_T \int_{-\infty}^{\infty} \exp\left(-\frac{\sqrt{p_T^2 + p_z^2 + m_0^2}}{T}\right) dp_z \tag{48}$$

$$f_y(y) \propto \cosh y \times \int_0^{p_{T\max}} p_T m_T \exp\left[-\frac{m_T \cosh y}{T}\right] dp_T, \tag{49}$$

$$f_{p_x}(p_x) \propto \int_{-\infty}^{\infty} \sqrt{p_x^2 + p_y^2 + m_0^2} \int_{y_{\min}}^{y_{\max}} \cosh y \times \exp\left[-\frac{\sqrt{p_x^2 + p_y^2 + m_0^2} \cosh y}{T}\right] dy dp_y$$

$$\text{or} \quad f_{p_x}(p_x) \propto \int_{-\infty}^{\infty}\int_{-\infty}^{\infty} \exp\left(-\frac{\sqrt{p_x^2 + p_y^2 + p_z^2 + m_0^2}}{T}\right) dp_z dp_y, \tag{50}$$

$$f_{p_y}(p_y) \propto \int_{-\infty}^{\infty} \sqrt{p_x^2 + p_y^2 + m_0^2} \int_{y_{\min}}^{y_{\max}} \cosh y \times \exp\left[-\frac{\sqrt{p_x^2 + p_y^2 + m_0^2} \cosh y}{T}\right] dy dp_x$$

$$\text{or} \quad f_{p_y}(p_y) \propto \int_{-\infty}^{\infty}\int_{-\infty}^{\infty} \exp\left(-\frac{\sqrt{p_x^2 + p_y^2 + p_z^2 + m_0^2}}{T}\right) dp_z dp_x, \tag{51}$$

$$f_{p_z}(p_z) \propto \int_0^{\infty} p_T \exp\left(-\frac{\sqrt{p_T^2 + p_z^2 + m_0^2}}{T}\right) dp_T. \tag{52}$$

To describe $p_T$ and $y$ spectra of particles produced in high energy collisions, we need usually to use a two-component or multi-component relativistic ideal gas model. In the two-component model, let $i$ denote the first or second component. According to Eq. (48), we have the $i$-th component in $p_T$ distribution to be

$$f_{p_T}(p_T) = A_i p_T m_T \int_{y_{\min}}^{y_{\max}} \cosh y \times \exp\left[-\frac{m_T \cosh y}{T_i}\right] dy, \tag{53}$$

where

$$A_i = 1\Big/ \int_0^{p_{T\max}} p_T m_T \int_{y_{\min}}^{y_{\max}} \cosh y \times \exp\left[-\frac{m_T \cosh y}{T_i}\right] dy dp_T$$

is the normalization constant. The $p_T$ distribution in the two-component model which is for the spectra in a given rapidity bin is

$$f_{p_T}(p_T) = kA_1 p_T m_T \int_{y_{\min}}^{y_{\max}} \cosh y \times \exp\left[-\frac{m_T \cosh y}{T_1}\right] dy$$
$$+ (1-k) A_2 p_T m_T \int_{y_{\min}}^{y_{\max}} \cosh y \times \exp\left[-\frac{m_T \cosh y}{T_2}\right] dy \quad (54)$$

where $T_1$ and $T_2$ denote the temperature parameters in the first and second components respectively, and $k$ denotes the contribution fraction of the first component. In Eqs. (53) and (54), the local source rapidity of each component is shifted to 0 to avoid the effect of directional movement.

According to Eq. (49), the $y$ distribution contributed by the $i$-th component with the source rapidity $Y_i$ is

$$f_y(y) = B_i \cosh(y - Y_i) \times \int_0^{p_{T\max}} p_T m_T \exp\left[-\frac{m_T \cosh(y - Y_i)}{T_i}\right] dp_T, \quad (55)$$

where

$$B_i = 1 \Big/ \int_{y_{\min}}^{y_{\max}} \cosh(y - Y_i) \times \int_0^{p_{T\max}} p_T m_T \exp\left[-\frac{m_T \cosh(y - Y_i)}{T_i}\right] dp_T dy$$

is the normalization constant. The $y$ distribution has a multi-component form due to each rapidity bin having a two-component distribution. In the framework of the overlapping cylinder model [19] in which the projectile cylinder and the target cylinder overlap around the mid-rapidity, we have the superposition of two continued-component forms in a wide rapidity range to be

$$f_y(y) = \frac{0.5}{\Delta Y + \delta Y} \left\{ \int_{-\Delta Y}^{\delta Y} B \cosh(y - Y) \times \int_0^{p_{T\max}} p_T m_T \exp\left[-\frac{m_T \cosh(y - Y)}{T}\right] dp_T dY \right.$$
$$\left. + \int_{-\delta Y}^{\Delta Y} B \cosh(y - Y) \times \int_0^{p_{T\max}} p_T m_T \exp\left[-\frac{m_T \cosh(y - Y)}{T}\right] dp_T dY \right\}, \quad (56)$$

where

$$B = 1 \Big/ \int_{y_{\min}}^{y_{\max}} \cosh(y - Y) \times \int_0^{p_{T\max}} p_T m_T \exp\left[-\frac{m_T \cosh(y - Y)}{T}\right] dp_T dy$$

is the normalization constant for the component with the local source rapidity $Y$, $T$ is the temperature which is insensitive to the distribution of $y$ and can be regarded as an invariant quantity, and $\Delta Y$ and $\delta Y$ ($\Delta Y \geq \delta Y \geq 0$) denote the rapidity shifts of the target and projectile cylinders. For a symmetrical collision system, the target cylinder stays in $[-\Delta Y, \delta Y]$ and the projectile cylinder stays in $[-\delta Y, \Delta Y]$ in the rapidity space. As a result of the overlapping cylinder model [19], Eq. (56) implies that a series of fireballs are assumed to distribute uniformly in the rapidity ranges $[-\Delta Y, \delta Y]$ and $[-\delta Y, \Delta Y]$.

## 4. Results and discussion

To show intuitively some distributions in terms of plot from the ideal gas model, as examples, we show some results in Figs. 1 and 2 by the analytical and Monte Carlo methods.

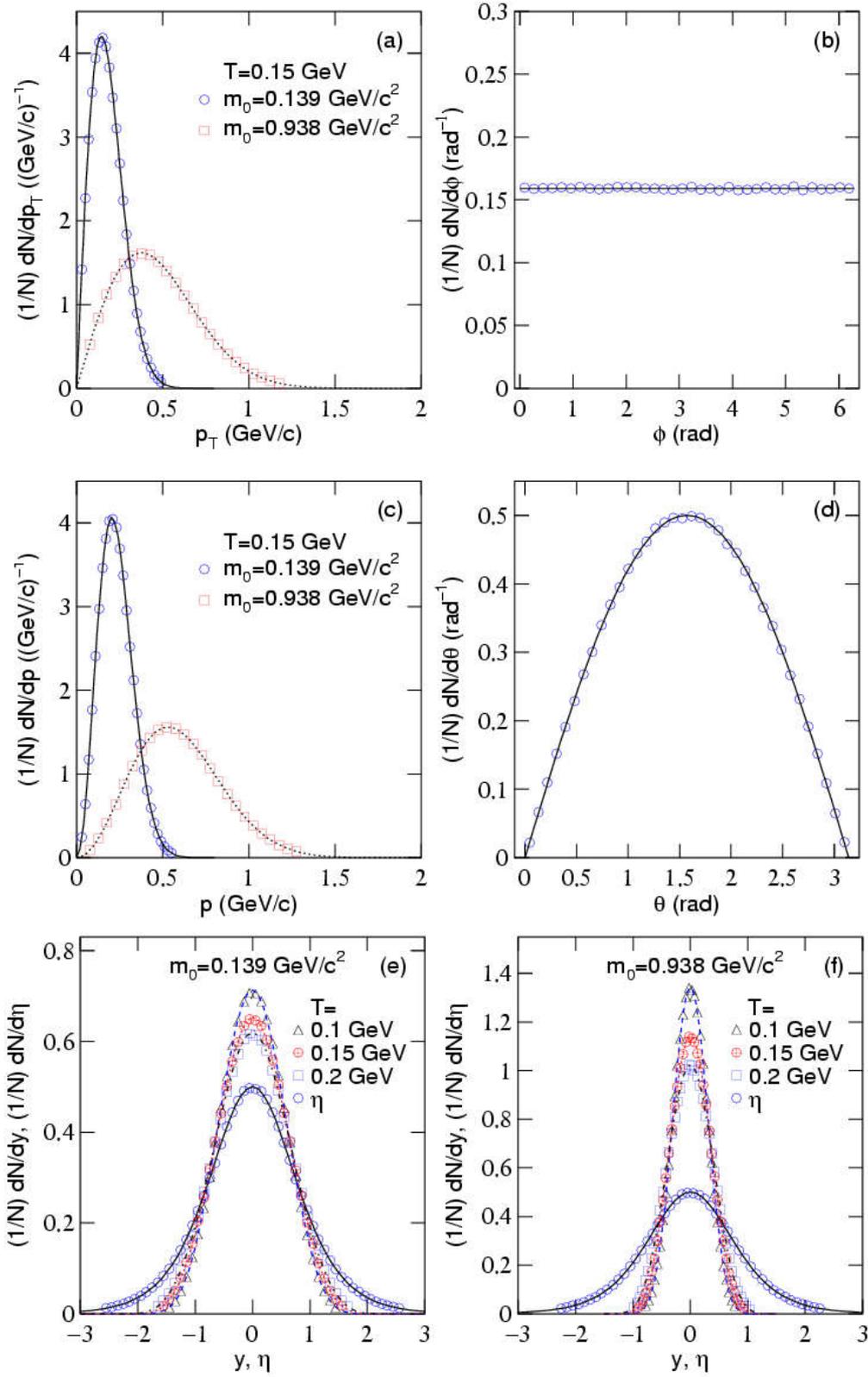

Fig. 1. Distributions of (a) $p_T$, (b) $\phi$, (c) $p$, and (d) $\theta$ for pions and protons, as well as (e) $y$ ($\eta$) for pions and (f) $y$ ($\eta$) for protons in the classical ideal gas model based on the distributions of momentum components, where the results for pions and protons in Fig. 1(b) [1(d)] are the same.

Figures 1(a)-1(d) present the distributions of $p_T$, $\phi$, $p$, and $\theta$, respectively, in the classical ideal gas model based on the distributions of $p_x$, $p_y$, and $p_z$. We have taken $m_0 = 0.139$ (pion) and $0.938$ GeV/$c^2$ (proton) as examples. The curves and symbols represent the results of analytic and Monte Carlo methods respectively, where the results for pions and protons in Fig. 1(b) [1(d)] are the same. Figures 1(e) and 1(f) present the distributions of $y$ at different $T$ shown in the panels for $m_0 = 0.139$ and $0.938$ GeV/$c^2$ respectively. As a comparison, the distribution of $\eta$ is also shown in the panels. One can see that the results of analytic and Monte Carlo methods are in agreement with each other. Although this observation is natural, the two results are confirmed each other and the methods used by us are confirmed to be correct. One can also see that the differences between the distributions of $y$ and $\eta$ are obvious. Although the differences for low mass particles at high temperature are small, we would rather use respectively the distributions of $y$ and $\eta$ for massive particles at any temperature. That is, we do not suggest that one of the two distributions is approximately replaced by another one.

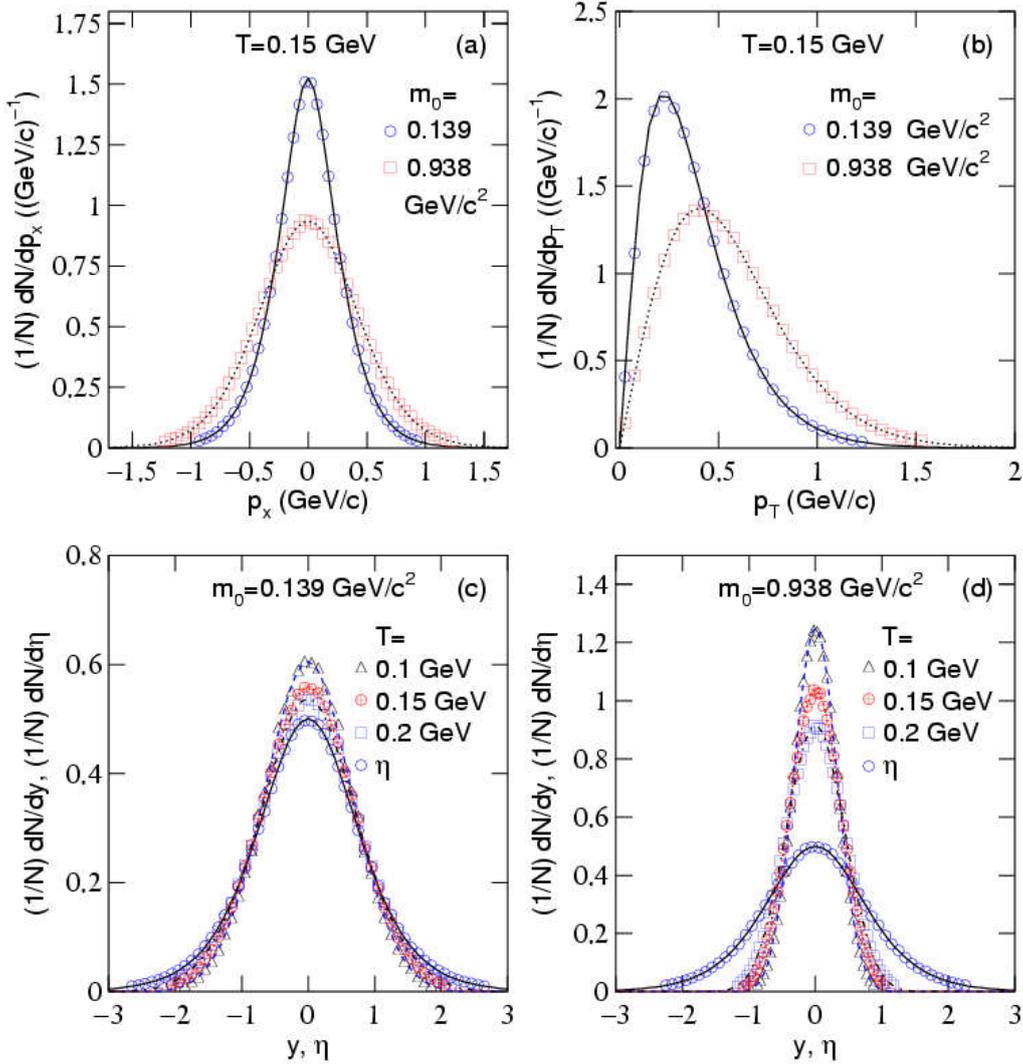

Fig. 2. Distributions of (a) $p_x$ and (b) $p_T$ for pions and protons, as well as (c) $y$ ($\eta$) for pions and (d) $y$ ($\eta$) for protons in the relativistic ideal gas model based on the distributions of momenta.

Figures 2(a) and 2(b) present the distributions of $p_x$ and $p_T$ respectively, in the relativistic ideal gas model based on the distributions of $p$ and isotropic assumption. We have taken $m_0 = 0.139$ and 0.938 GeV/$c^2$ as examples. The curves and symbols represent the results of analytic and Monte Carlo methods respectively. Figures 2(c) and 2(d) present the distributions of $y$ at different $T$ shown in the panels for $m_0 = 0.139$ and 0.938 GeV/$c^2$ respectively. As a comparison, the distribution of $\eta$ is also shown in the panels. Once again, one can see that the results of analytic and Monte Carlo methods are in agreement with each other. This observation is natural, which confirms that the two results obtained by us and the methods used by us are correct. The conclusions on the distributions of $y$ and $\eta$ obtained from Fig. 1 is also obtained from Fig. 2. Comparing with that in the classical ideal gas model, the distribution of $y$ in the relativistic ideal gas model is closer to the distribution of $\eta$ at a given temperature.

In particular, if we use the distributions of $p_T$ and/or $y$ in the relativistic ideal gas model to "measure" (fit) those in the classical one, a lower temperature will be obtained. Contrarily, if we use the distributions of $p_T$ and/or $y$ in the classical ideal gas model to "measure" those in the relativistic one, a higher temperature will be obtained. In other words, the temperature of thermodynamic system decreases by dividing the Lorentz factor ($\gamma$) in relativistic situation. This observation confirms our previous work [20], which shows that the Planck-Einstein relation [21-24] is right. A moving system which has no energy current exchange with external surroundings becomes cool.

To show clearly relativistic temperature transformation, let $T_0$ denote the temperature of the system at rest frame, $T_f$ denote temperature of the system when it moves with $\gamma$. We have $T_f = T_0/\gamma < T_0$ ($\gamma > 1$) which is the Planck-Einstein relation. This formula can be understood by the Lorentz constriction of the system size in the moving direction, which limits the moving space of the ideal gas particles and results in smaller volume of the system. In the case of considering isobaric process [25], one can obtain a lower temperature of the system. In addition, there are other relations such as $T_f = T_0\gamma > T_0$ and $T_f = T_0$. A recent review article [25] had not made a conclusion on the relative size of $T_f$ and $T_0$. We insist on the Planck-Einstein relation to be correct due to invariant $p_T$ spectra in high energy collisions [20].

We would like to point out that the temperature discussed above is the so-called effective temperature. It is not the "real" temperature of emission source at the kinetic freeze-out in high energy collisions. To extract the "real" temperature at the kinetic freeze-out, the influence of flow effect should be excluded. After excluding the flow effect, the "real" temperature will be smaller than the effective temperature. As for how to extract the effective temperature is beyond the focus of the present work. We shall not discuss this issue further.

As an application of the relativistic ideal gas model in the studies of particle productions in high energy collisions, we analyze the $p_T$ and $y$ spectra of negative pions produced in proton-proton (p-p) collisions at center-of-mass energy $\sqrt{s} = 6.3$, 7.7, 8.8, 12.3, and 17.3 GeV in Figs. 3(a) and 3(b) respectively. The symbols in Fig. 3(a) represent the experimental data of the NA61/SHINE Collaboration [26] measured in the rapidity range $y = 0 - 0.2$. The closed symbols in Fig. 3(b) represent the experimental data of the NA61/SHINE Collaboration [26] and the open ones are reflected at the mid-rapidity $y = 0$. The curves for the $p_T$ and $y$ spectra are our results fitted by Eqs.

(54) and (56) respectively. The corresponding parameters with $\chi^2$ per degree of freedom (dof) are listed in Tables 1 and 2. One can see that the two-component relativistic ideal gas model describes the $p_T$ spectra of negative pions produced in the given rapidity bin in p-p collisions at the mentioned energies. When we use the model to describe the $y$ spectra, a series of emission sources stayed in two overlapping rapidity ranges have to be considered synchronously. This results in the multi-component relativistic ideal gas model.

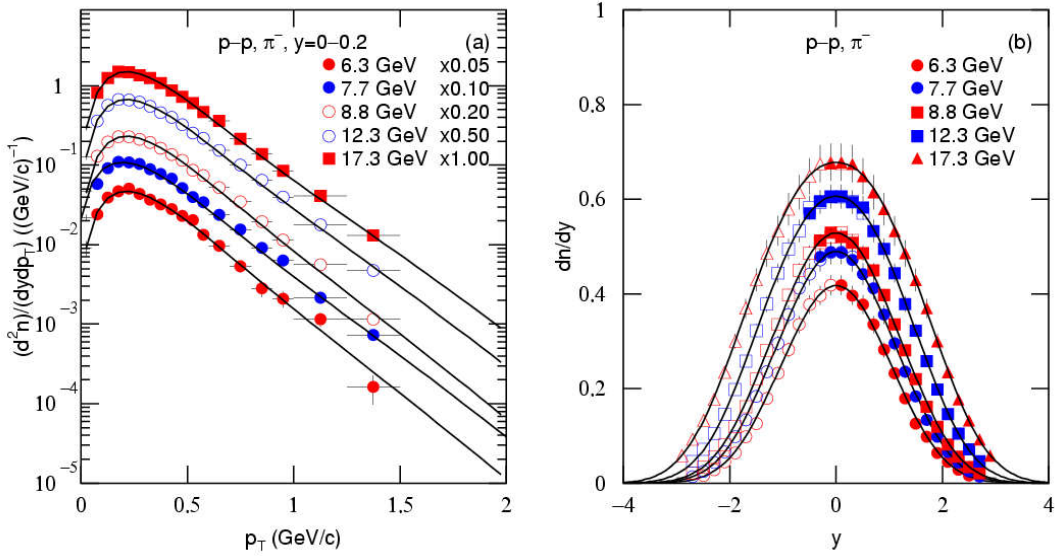

Fig. 3. The spectra of (a) $p_T$ and (b) $y$ of negative pions produced in (a) mid-rapidity range 0−0.2 and (b) wide rapidity range in p-p collisions at $\sqrt{s} = 6.3$, 7.7, 8.8, 12.3, and 17.3 GeV. The solid symbols represent the data measured by the NA61/SHINE Collaboration [26] and the open ones are reflected at the mid-rapidity $y = 0$. The curves for the $p_T$ and $y$ spectra are our results fitted by Eqs. (54) and (56) respectively. The spectra in Fig. 3(a) are scaled by different amounts marked in the panel for clearly.

It should be noted that we only applied the distributions to the p-p data in Fig. 3. However, the situation is similar when the distributions are applied to the data obtained in hadron-nucleus and nucleus-nucleus collisions. In fact, the data from p-p, hadron-nucleus, and nucleus-nucleus collisions demonstrate similarity in hadron-production processes which are widely discussed in [27-37] in terms of similarity. Not only for the shapes of $p_T$ or $y$ spectrum at given energy, but also for the variant trends of parameters in $p_T$ or $y$ spectrum on collision energy, the similarity is widely existed in high energy collisions.

Although there are many modern models being used in the analyses of particle production in high energy collisions, the relativistic ideal gas model and its multi-component form have still strong vitality. In our opinion, the relativistic ideal gas is not out-of-date. It is not strange that the Fermi-Dirac and Bose-Einstein statistics which are widely used in modern physics are based on the relativistic ideal gas model, though the chemical potential and spin property are included in the two statistics. The multi-component relativistic ideal gas model is in fact the customary form of the thermalized cylinder [19] or multi-source thermal model [38] in which the single model can be replaced by other models and distributions.

Table 1. The parameter values and $\chi^2/\mathrm{dof}$ corresponding to the curves in Fig. 3(a).

| $\sqrt{s_{NN}}$ (GeV) | $T_1$ (GeV) | $T_2$ (GeV) | $k$ | $N_0$ | $\chi^2/\mathrm{dof}$ |
|---|---|---|---|---|---|
| 6.3 | 0.100±0.005 | 0.166±0.005 | 0.78±0.02 | 0.082±0.006 | 68/14 |
| 7.7 | 0.100±0.003 | 0.170±0.006 | 0.77±0.03 | 0.096±0.008 | 23/14 |
| 8.8 | 0.100±0.006 | 0.167±0.003 | 0.76±0.03 | 0.102±0.006 | 37/14 |
| 12.3 | 0.100±0.006 | 0.175±0.005 | 0.75±0.03 | 0.120±0.008 | 24/14 |
| 17.3 | 0.100±0.007 | 0.180±0.005 | 0.76±0.03 | 0.134±0.008 | 19/14 |

Table 2. The parameter values and $\chi^2/\mathrm{dof}$ corresponding to the curves in Fig. 3(b).

| $\sqrt{s_{NN}}$ (GeV) | $T$ (GeV) | $\Delta Y$ | $\delta Y$ | $N_0$ | $\chi^2/\mathrm{dof}$ |
|---|---|---|---|---|---|
| 6.3 | 0.115±0.005 | 1.35±0.04 | 0.48±0.09 | 1.047±0.043 | 8/9 |
| 7.7 | 0.116±0.004 | 1.50±0.06 | 0.55±0.15 | 1.305±0.035 | 1/12 |
| 8.8 | 0.114±0.005 | 1.62±0.05 | 0.68±0.14 | 1.478±0.023 | 3/13 |
| 12.3 | 0.119±0.006 | 1.87±0.05 | 1.03±0.10 | 1.920±0.032 | 1/13 |
| 17.3 | 0.119±0.007 | 2.15±0.03 | 1.20±0.10 | 2.408±0.032 | 1/12 |

## 5. Conclusions

The mutual derivation between distributions of momenta and momentum components in high energy collisions were systematically studied in this paper. The results of analytic and Monte Carlo methods are in agreement with each other. Not only the two results are confirmed each other, but also the two methods used by us are confirmed to be correct.

The distributions of rapidities and pseudorapidities were studied. The classical and relativistic ideal gas models were used as examples to show some distributions by the analytic and Monte Carlo methods. The differences between the distributions of rapidities and pseudorapidities for low mass particles at high temperature are small, though we do not suggest replace the opposite side each other. Comparing with that in the classical ideal gas model, the distribution of rapidities in the relativistic ideal gas model is closer to the distribution of pseudorapidities at a given temperature.

From the distributions of transverse momenta and/or rapidities, the temperature parameter can be obtained. When comparing with that in the classical situation, the temperature of thermodynamic system in relativistic situation decreases by dividing the Lorentz factor. This confirms the Planck-Einstein relation which indicates that a moving system which has no energy current exchange with external surroundings becomes cool.

The relativistic ideal gas model and its multi-component form have still strong vitality in high energy collisions. In the transverse plane such as in the transverse momentum space, two-component form can describe the transverse momentum spectra of light flavor particles in the given rapidity bin. In the longitudinal direction such as in the rapidity space, multiple sources which result in the multi-component form such as the overlapping cylinder model are needed to describe the rapidity spectra.

**Data Availability**

The data used to support the findings of this study are quoted from the mentioned references. As a phenomenological work, this paper does not report new data.

**Conflicts of Interest**

The authors declare that there are no conflicts of interest regarding the publication of this paper.

**Acknowledgments**

This work was supported by the National Natural Science Foundation of China under Grant Nos. 11575103 and 11847311, the Shanxi Provincial Natural Science Foundation under Grant No. 201701D121005, and the Fund for Shanxi "1331 Project" Key Subjects Construction.